\documentclass{elsart}
\usepackage{epsfig}
\def\TWIST{$\mathcal{TWIST}\ $}
\def\Li#1#2{{\mathrm{Li}}_{#1}\left(#2\right)}
\def\SSS#1#2{{\mathrm{S}}_{#1}\left(#2\right)}
\def\Sot#1{{\mathrm{S}}_{1,2}\left(#1\right)}
\def\ba{\begin{eqnarray}}
\def\ea{\end{eqnarray}}
\def\dd{{\mathrm d}}

\def\ga{\mathrel{\mathpalette\fun >}}
\def\fun#1#2{\lower3.6pt\vbox{\baselineskip0pt\lineskip.9pt
  \ialign{$\mathsurround=0pt#1\hfil##\hfil$\crcr#2\crcr\sim\crcr}}}
\def\order#1{{\mathcal O}\left(#1\right)}
\newcommand{\vecc}[1]{\mbox{\boldmath $#1$}}

\begin{document}
\begin{frontmatter}

\title{Virtual and Soft Pair Corrections to Polarized Muon Decay Spectrum}

\author{A.B. Arbuzov\thanksref{UofA}}

\thanks[UofA]{A certain part of this work was
performed in University of Alberta, Edmonton, Canada}

\address{Bogoliubov Laboratory of Theoretical Physics, \\
JINR,\ Dubna, \ 141980 \ \  Russia \\
{\tt e-mail:} arbuzov@thsun1.jinr.ru
}

\begin{abstract}
Radiative corrections to the muon decay spectrum due to soft
and virtual electron--positron pairs are calculated.
\end{abstract}


\end{frontmatter}

\section{Introduction}

The experiment \TWIST\cite{Rodning:2001js,Quraan:2000vq}
is currently running at Canada's National Laboratory TRIUMF.
It is going to measure the muon decay
spectrum~\cite{Fetscher:2000th,Kuno:2001jp}
with the accuracy level of about $1\cdot 10^{-4}$.
That will make a serious test of the space--time structure
of the weak interaction. The experiment is able to put stringent limits
on a bunch of parameters in models beyond the Standard Model (SM),
{\it e.g.}, on the mass and the mixing angle of a possible right--handed $W$-boson.
To confront the experimental results with the SM,
adequately accurate theoretical predictions should
be provided. This requires to calculate radiative corrections within the perturbative
Quantum Electrodynamics (QED). Here we will present analytical results for
two specific contributions related to radiation of virtual and soft real
electron--positron pairs. The corrections under consideration are of the
order $\order{\alpha^2}$, where $\alpha$ is the fine structure constant.

The contributions of virtual $\mu^+\mu^-$, $\tau^+\tau^-$, and hadronic pairs
were found~\cite{Davydychev:2001ee} to be small compared with the $1\cdot 10^{-4}$
precision tag of the modern experiments. The contribution of $e^+e^-$ pairs is
enhanced by powers of the large logarithm $L=\ln(m_\mu^2/m_e^2)\approx 10.66$.
Analysis of the leading and next--to--leading terms from this correction
in Refs.~\cite{Arbuzov:2002rp,Arbuzov:2002cn} has shown that the numerical effect is not as
small as for other leptonic flavors, and it should be taken into account.
Comparison of the leading and next--to--leading contributions revealed a poor
convergence of the series in $L$. Calculation of the terms without the large
logarithm was found to be desirable.

Within the Standard Model, the differential
distribution of electrons (summed over electron spin states)
in the polarized muon decay can be represented as
\ba \label{general}
&& \frac{\dd^2\Gamma^{\mu^{\mp}\to
e^{\mp}\nu\bar{\nu}}}{\dd x\dd c}
= \Gamma_0 \left[ F(x) \pm cP_{\mu} G(x) \right], \qquad
\Gamma_0 = \frac{G_F^2 m_\mu^5}{192\pi^3}\, ,
\nonumber \\ &&
c = \cos\theta, \qquad
x = \frac{2m_{\mu}E_e}{m_\mu^2+m_e^2}, \qquad
x_0 \leq x \leq 1, \qquad
x_0 = \frac{2m_{\mu}m_e}{m_\mu^2+m_e^2},
\ea
where
$m_\mu$ and $m_e$ are the muon and electron masses;
$G_F$ is the Fermi coupling constant;
$\theta$ is the angle between the muon polarization vector $\vec{P}_{\mu}$
and the electron (or positron) momentum;
$E_e$ and $x$ are the energy and the energy fraction of $e^{\pm}$.
Here we adopt the definition of the Fermi coupling constant following
Ref.~\cite{vanRitbergen:2000fi}.
Functions $F(x)$ and $G(x)$ describe the isotropic and anisotropic
parts of the spectrum, respectively. Within perturbative QED, they
can be expanded in series in $\alpha$:
\ba
F(x) = f_{\mathrm{Born}}(x) + \frac{\alpha}{2\pi}f_1(x)
+ \biggl(\frac{\alpha}{2\pi}\biggr)^2f_2(x)
+ \biggl(\frac{\alpha}{2\pi}\biggr)^3f_3(x)
+ \order{\alpha^4},
\ea
and in the same way for $G(x)$.
Among different contributions into the functions $F(x)$ and $G(x)$
(see Ref.~\cite{Arbuzov:2002rp} for details),
there are ones related to the production of electron--positron pairs.
In this Letter we will consider the effect of soft and virtual $e^+e^-$
pairs.

\section{Soft $e^+e^-$ Pairs}

The process of real pair production doesn't reveal any infrared
singularity, contrary to the case of photon radiation. Nevertheless,
a separate consideration of soft pair emission can be of interest.
In fact, $e^+e^-$ pairs with energy below a certain threshold can't
be observed in experiments with muons decaying at rest. So, the corresponding
contribution is a specific correction to the measured decay spectrum.
Moreover, the behavior of the real pair emission in the soft limit is not
smooth. An integration over the domain between the threshold of real pair
production and a certain cut on the maximal energy of the soft pair
is desirable.

The maximal energy of the soft pair is defined by the parameter $\Delta$,
which is assumed to be large compared with the electron mass:
\ba \label{Esoft}
E_{\mathrm{pair}} \leq \Delta\frac{m_\mu}{2}, \qquad
\frac{m_e}{m_\mu} \ll \Delta \ll 1.
\ea

Due to the smallness of the pair component energies, the matrix element
$M$ of the process
\ba
\mu^-(p) \ \longrightarrow e^-(q)\ +\ \nu_{\mu}(r_1)\ +\ \bar{\nu}_e(r_2)\
+\ e^+(p_+)\ +\ e^-(p_-)
\ea
can be expressed as a product
of the matrix element $M_0$ of the hard sub--process (the non--radiative
muon decay) and the classic accompanying radiation factor:
\ba
M = M_0 \frac{4\pi\alpha}{k^2}\bar{v}(p_+)\gamma^{\mu}u(p_-)J_{\mu},
\qquad k = p_+ + p_-,
\ea
where $p_{+,-}$ are the momenta of the positron and electron from
the created pair.
The radiation factor reads
\ba
J_{\mu} = \frac{p_{\mu}}{pk - \frac{1}{2}k^2}
- \frac{q_{\mu}}{qk + \frac{1}{2}k^2}\, .
\ea

Performing the covariant integration of the summed over spin states
modulus of the matrix element over the pair components momenta, we obtain
\begin{eqnarray}
&& \sum\limits_{\mbox{spin}}|\bar{v}(p_+)\gamma^{\mu}u(p_-)|^2
= 4(p_+^{\mu}p_-^{\nu}+p_+^{\nu}p_-^{\mu}-\frac{k^2}{2}\; g^{\mu\nu}),
\nonumber \\ \nonumber
&& \int\frac{\dd^3\vecc{p}_+ \dd^3\vecc{p}_-}{p_+^0 p_-^0}
\delta^{4}(p_+ + p_- - k)
(p_+^{\mu} p_-^{\nu} + p_+^{\nu} p_-^{\mu} - \frac{k^2 }{2}g^{\mu\nu}) =
\\ && \qquad
= \biggl(-\frac{2\pi}{3}(k^2+2m_e^2)\sqrt{1-\frac{4m_e^2}{k^2}}\;\biggr)
(g^{\mu\nu}-\frac{1}{k^2} k^{\mu}k^{\nu}).
\end{eqnarray}
It is convenient to parameterize the phase volume of
the total pair momentum as
\begin{eqnarray}
\dd^4k=\dd k_0 \vecc{k}^2 \dd |\vecc{k}| \dd\Omega_k
      =\pi \dd k_0 \dd k^2 \, \sqrt{k_0^2-k^2}\; \dd c_k\dd\varphi_k\, ,
\end{eqnarray}
where a trivial integration over the azimuthal angle can be performed:
$\int\dd\varphi_k \to 2\pi$.
Now I integrate over the total pair momentum with the condition~(\ref{Esoft})
$(k_0 \equiv E_{\mathrm{pair}})$.
In this way I got the following result for the soft pair contribution:
\ba \label{SP}
&& \frac{\dd\Gamma^{\mathrm{SP}}}{\dd c\; \dd x} =
\frac{\dd\Gamma^{\mathrm{Born}}}{\dd c\; \dd x} \delta^{\mathrm{SP}}, \qquad
\frac{\dd\Gamma^{\mathrm{Born}}}{\dd c\; \dd x} = \Gamma_0\left[f_0(x)
\pm cP_{\mu} g_0(x) \right] + \order{\frac{m_e^2}{m_{\mu}^2}},
\nonumber \\
&& f_0(x) = x^2(3-2x), \qquad g_0(x) = x^2(1-2x),
\nonumber \\
&& \delta^{\mathrm{SP}} = \frac{\alpha^2}{3\pi^2}\biggl[
\frac{1}{12}\ln^3A - \frac{2}{3}\ln^2A
+ \ln A\biggl(\frac{61}{18}-\zeta(2)\biggr)
- \frac{223}{27}
\nonumber \\ && \qquad
+ \frac{8}{3}\zeta(2) + 2\zeta(3)
\biggr], \\ \nonumber
&& \ln A = L+2\ln\Delta, \qquad
\zeta(n) = \sum_{k=1}^{\infty}\frac{1}{k^n}, \qquad
\zeta(2) = \frac{\pi^2}{6}\, .
\ea
So we calculated explicitly all the terms in $\delta^{\mathrm{SP}}$
except the ones suppressed
by the small factors $(\alpha/\pi)^2m_{e}^2/m_{\mu}^2$
and $(\alpha/\pi)^2\Delta$.

\section{Virtual $e^+e^-$ Pair}

We will use here the substitution suggested by J.~Schwinger for the photon
propagator (with 4--momentum $k$) corrected by a one--loop vacuum
polarization insertion:
\ba \label{schwinger}
&& \frac{1}{k^2-\lambda^2+i0}\rightarrow\frac{\alpha}{\pi}\int\limits_{0}^{1}
\frac{\dd v\phi(v)}{1-v^2}\,\, \frac{1}{k^2-M^2+i0}\, , \quad
M^2=\frac{4m_2^2}{1-v^2}\, , \\ \nonumber && \quad
\phi(v)=\frac{2}{3} - \frac{1}{3}(1-v^2)(2-v^2),
\ea
where $m_2$ is the mass of the fermion in the loop.

The standard technique of integration over Feynman parameters
can be used here. We are interested in the region of electron energy
fractions $z\gg m_e/m_\mu$. Analytical expressions for the relevant
integrals in this region are given in Appendix A. As concerning the
region of small electron energy fractions $(z\sim m_e/m_\mu)$,
it requires a more
accurate treatment. But the differential width there is rapidly
decreasing (see {\it i.e.} the Born--level functions in Eq.~(\ref{SP})),
and the contribution of this region into the total width is
suppressed by the mass ratio squared.

Formally, we have an ultraviolet singularity in the virtual pair
correction. The Fermi theory is not renormalizable in the general case.
But for the muon decay everything is safe, since the standard
renormalization of the electron and muon wave functions removes
the singularity~\cite{berman:1962xx}. Note that we need to
use here only the pair contribution into the renormalization constants.
They can be found easily from the calculation of the virtual pair
corrections to the $ee$ and $\mu\mu$ vertexes (see Appendix B),
where we had
\ba
&& \ln\frac{\Lambda^2}{D_e} \longrightarrow - 2V_e(0), \qquad
\ln\frac{\Lambda^2}{D_\mu} \longrightarrow - 2V_\mu(0),
\nonumber \\ &&
V_\mu(0) =  - \frac{1}{12}L^2 + \frac{25}{36}L
- \frac{325}{216} - \frac{1}{3}\zeta(2), \qquad
V_e(0) = - \frac{469}{216} + \frac{4}{3}\zeta(2),
\ea
where $\Lambda$ is the ultraviolet cut--off;
see Appendix~A for quantities $D_{e,\mu}$.
For the muon decay we use {\em half a sum\/} of the above
substitutions:
\ba
&& \frac{1}{2}\ln\frac{\Lambda^2}{D_\mu}
+ \frac{1}{2}\ln\frac{\Lambda^2}{D_e}
\longrightarrow - V_\mu(0) - V_e(0).
\ea
The same logic was applied for
renormalization of one--loop corrections
to muon decay~\cite{Arbuzov:2001ui}.

I got the following result for the virtual $e^+e^-$ pair contribution:
\ba
\frac{\Gamma^{\mathrm{VP}}}{\dd c\;\dd x} =
\Gamma_0 \biggl(\frac{\alpha}{2\pi}\biggr)^2 \biggl[
f_{2,\mathrm{virt}}^{(e^+e^-)}(x)
\pm cP_{\mu} g_{2,\mathrm{virt}}^{(e^+e^-)}(x)
+ \order{\frac{m_e^2}{m_\mu^2}} \biggr],
\ea
where
\ba \label{virtp}
&& f_{2,\mathrm{virt}}^{(e^+e^-)}(x) = f_0(x) W(x)
- 2x^2\ln x L - 2x^2\ln^2x - 2x^2\Li{2}{1-x}
\nonumber \\ && \qquad
+ 7x^2\ln x + \frac{2}{3}x\ln x
+ \frac{2}{3}\ln x - \frac{2}{3(1-x)}\ln x,
\nonumber \\ \nonumber
&& g_{2,\mathrm{virt}}^{(e^+e^-)}(x) = g_0(x) W(x)
- \frac{2}{3}x^2\ln x L
- \frac{2}{3}x^2\biggl( \Li{2}{1-x} + \ln^2x \biggr)
\\ \nonumber && \qquad
+ \frac{13}{9}x^2\ln x
- \frac{2}{3}x\ln x - \frac{2}{3}\ln x + \frac{2}{3(1-x)}\ln x,
\\ \nonumber
&& W(x) = - \frac{1}{9}L^3
+ \biggl( \frac{25}{18} - \frac{2}{3}\ln x \biggr)L^2
+ \biggl( - \frac{4}{3}\Li{2}{1-x} - \frac{4}{3}\ln^2x
\nonumber \\ && \qquad
+ \frac{38}{9}\ln x - \frac{4}{3}\zeta(2) - \frac{397}{54} \biggr)L
- \frac{8}{3}\Sot{1-x} + \frac{4}{3}\Li{3}{1-x}
+ \frac{38}{9}\Li{2}{1-x}
\nonumber \\ && \qquad
- \frac{8}{9}\ln^3x - \frac{8}{3}\ln x \Li{2}{1-x}
+ \frac{38}{9}\ln^2x - \frac{8}{3}\zeta(2)\ln x
- \frac{265}{27}\ln x
\nonumber \\ && \qquad
+ \frac{4}{3}\zeta(3) + \frac{22}{9}\zeta(2) + \frac{517}{27},
\\ \nonumber &&
\Li{2}{x} \equiv - \int\limits_{0}^{x}\dd y\;\frac{\ln(1-y)}{y}\, ,\qquad
\Li{3}{x} \equiv \int\limits_{0}^{x}\dd y\;\frac{\Li{2}{y}}{y}\, ,
\\ \nonumber &&
\Sot{x} \equiv \frac{1}{2}\int\limits_{0}^{x}\dd y\;\frac{\ln^2(1-y)}{y}\, .
\ea
It is worth to note that the sub--leading virtual corrections don't factorize
before the Born functions $f_0(x)$ and $g_0(x)$.

By integration over the energy fraction and the angle we receive
the corresponding contribution to the total muon width:
\ba \label{gvp}
\Gamma^{\mathrm{VP}} &=& \int\limits_{-1}^{1}\dd c
\int\limits_{0}^{1}\dd x \; \frac{\Gamma^{\mathrm{VP}}}{\dd c\;\dd x}
= \Gamma_0\biggl(\frac{\alpha}{2\pi}\biggr)^2 \biggl[
- \frac{1}{9}L^3 + \frac{5}{3}L^2 - \biggl( \frac{265}{36}
+ \frac{8}{3}\zeta(2) \biggr)L
\nonumber \\
&+& \frac{20063}{1296}
+ \frac{61}{9}\zeta(2) + \frac{16}{3}\zeta(3) \biggr]
\approx - 5.0497\cdot 10^{-5}\; \Gamma_0.
\ea

This quantity was calculated earlier in Ref.~\cite{vanRitbergen:1998yd}
by numerical integration using dispersion relations:
\ba
\Gamma^{\mathrm{VP}}([10]) \approx - 5.1326 \cdot 10^{-5}\; \Gamma_0,
\ea
which is close but different from my number. The reason for this
discrepancy will be investigated elsewhere. At least part of the difference
can be due to terms proportional to $(\alpha/\pi)^2(m_e^2/m_\mu^2)L^n$,
which were omitted in my calculation.

The correction to the forward--backward asymmetry of the decay
can be found also:
\ba
\Gamma^{\mathrm{VP}}_{\mathrm{FB}} &=& \left[\int\limits_{0}^{1}\dd c
- \int\limits_{-1}^{0}\dd c \right]
\int\limits_{0}^{1}\dd x \; \frac{\Gamma^{\mathrm{VP}}}{\dd c\;\dd x}
= \Gamma_0\biggl(\frac{\alpha}{2\pi}\biggr)^2 \biggl[
\frac{1}{54}L^3 - \frac{13}{54}L^2
+ \biggl( \frac{647}{648}
\nonumber \\
&+& \frac{4}{9}\zeta(2) \biggr)L
- \frac{10339}{7776} - \frac{3}{2}\zeta(2)
- \frac{8}{9}\zeta(3) \biggr]
\approx - 1.17\cdot 10^{-5}\; \Gamma_0.
\ea

\section{Numerical Results and Conclusions}

\begin{figure}[ht]
\epsfig{file=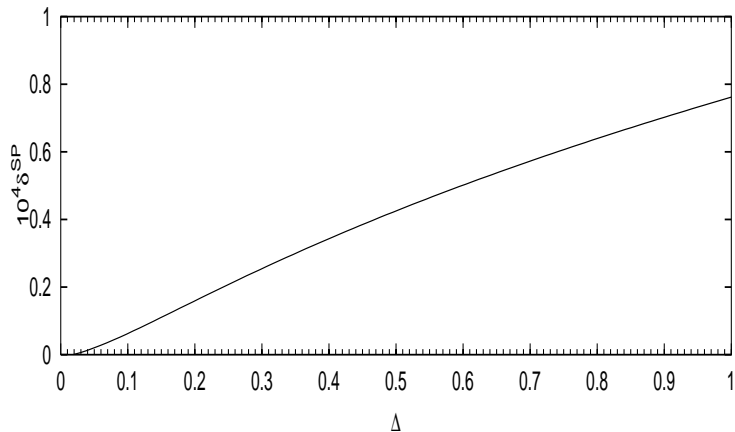,width=10cm,height=6cm}
\caption{The relative effect of soft pair corrections {\it versus}
the cut value.}
\label{figure:1}
\end{figure}

The relative effect of the soft pair correction depends only on the cut value. It
is shown in Fig.~\ref{figure:1}.
The soft pair approximation~(\ref{Esoft}) is not valid for values of $\Delta$ close to
the threshold of real pair production and for large $\Delta \sim 1$. But it can
be used there as a simple estimate. So, by taking $\Delta=1$ we make an estimate of
the order of magnitude of the total contribution due to real $e^+e^-$ pairs
(here the estimate is about two times the true value).
For very small values of $\Delta$ the correction should vanish in any case, so the
approximation is really safe there.

\begin{figure}[ht]
\epsfig{file=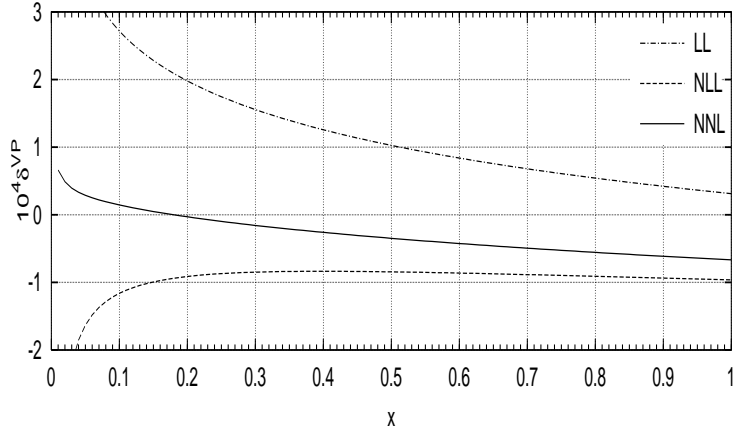,width=10cm,height=6cm}
\caption{The relative effect of virtual pair corrections {\it versus}
electron energy fraction in different approximations.}
\label{figure:2}
\end{figure}

Let us define the relative contribution of the virtual $e^+e^-$ pair corrections
in the form
\ba
\delta^{\mathrm{VP}}(x) = \left(\frac{\alpha}{2\pi}\right)^2
\frac{f_{2,\mathrm{virt}}^{(e^+e^-)}(x)
+ cP_{\mu}g_{2,\mathrm{virt}}^{(e^+e^-)}(x)}
{f_0(x) + cP_{\mu}g_0(x)}\, .
\ea
The dependence of this function on the
electron energy fraction is shown in Fig.~\ref{figure:2}
in different approximations for $P_{\mu}=1$, $c=1$.
The dependence on $c$ is very weak, because the main
part of the correction is factorized before the Born--level functions
and cancels out in the ratio.
The leading logarithmic (LL) approximation takes into account only the
terms of the order $\order{\alpha^2L^3,\alpha^2L^2}$,
the next--to--leading logarithmic (NLL)
approximation includes also the $\order{\alpha^2L^{1}}$ terms, and the
next--to--next--to--leading approximation (NNL) represents the complete result.

The third power of the large logarithm cancels out in the sum of the virtual
and soft pair contributions:
\ba
\frac{\Gamma^{\mathrm{SVP}}}{\dd c\;\dd x} =
\Gamma_0 \biggl(\frac{\alpha}{2\pi}\biggr)^2 \biggl[
f_{2,\mathrm{SV}}^{(e^+e^-)}(x)
\pm cP_{\mu} g_{2,\mathrm{SV}}^{(e^+e^-)}(x)
+ \order{\frac{m_e^2}{m_\mu^2},\,\Delta} \biggr],
\ea
where
\ba
f_{2,\mathrm{SV}}^{(e^+e^-)}(x) &=& f_0(x) U(x)
- 2x^2\ln x L - 2x^2\ln^2x - 2x^2\Li{2}{1-x}
\nonumber \\
&-& \frac{2}{3(1-x)}\ln x + \frac{2}{3}x\ln x
+ 7x^2\ln x  + \frac{2}{3}\ln x,
\nonumber \\
g_{2,\mathrm{SV}}^{(e^+e^-)}(x) &=& g_0(x) U(x)
- \frac{2}{3}x^2\ln x L
- \frac{2}{3}x^2\ln^2x - \frac{2}{3}x^2\Li{2}{1-x}
\nonumber \\
&+& \frac{2}{3(1-x)}\ln x
- \frac{2}{3}x\ln x + \frac{13}{9}x^2\ln x  - \frac{2}{3}\ln x,
\nonumber \\
U(x) &=& \biggl( \frac{1}{2} + \frac{2}{3}\ln\Delta - \frac{2}{3}\ln x \biggr)L^2
+ \biggl( \frac{4}{3}\ln^2\Delta - \frac{32}{9}\ln\Delta
\nonumber \\
&-& \frac{4}{3}\Li{2}{1-x}
- \frac{4}{3}\ln^2x + \frac{38}{9}\ln x
- \frac{17}{6} - \frac{8}{3}\zeta(2) \biggr)L
\nonumber \\
&+& \frac{8}{9}\ln^3\Delta - \frac{32}{9}\ln^2\Delta
- \frac{8}{3}\zeta(2)\ln\Delta + \frac{244}{27}\ln\Delta
\nonumber \\
&+& \frac{4}{3}\Li{3}{1-x} - \frac{8}{3}\Sot{1-x} - \frac{8}{9}\ln^3x
- \frac{8}{3}\ln x \Li{2}{1-x}
\nonumber \\
&+& \frac{38}{9}\Li{2}{1-x} + \frac{38}{9}\ln^2x
- \frac{8}{3}\zeta(2)\ln x - \frac{265}{27}\ln x
\nonumber \\
&+& \frac{659}{81} + 6\zeta(2) + 4\zeta(3).
\ea
I checked that the leading and next--to--leading terms in the above formula
agree with the corresponding contribution obtained within the fragmentation
function formalism in Refs.~\cite{Arbuzov:2002rp,Arbuzov:2002cn}.

In this way we simulate the experimental set--up
with a certain energy threshold for registration of pairs, while events
with pair production above the threshold (with several visible charged
particles in the final state) are rejected.

If the radiation of real pairs is completely forbidden by kinematics
(or experimental conditions),
only the virtual corrections~(\ref{virtp}) contribute.
That happens, for instance at large values of $x\ga 0.99$.

Thus, two contributions to the total set of radiative corrections
for the muon decay spectrum are presented. They are required to reach 
the level of the theoretical accuracy below $1\cdot10^{-4}$. The formulae
can be used for semi--analytical estimates and as a part of a Monte Carlo
code to describe the pair production contribution to the decay spectrum.
The formulae are valid also for pair corrections to leptonic $\tau$-decays.  

\ack{
I am grateful to V. Bytev and E. Kuraev for discussions.
This work was supported by RFBR grant 03-02-17077.
}

\section*{Appendix A\\
List of integrals for the virtual pair correction}

\setcounter{equation}{0}
\renewcommand{\theequation}{A.\arabic{equation}}

Here we give the list of integrals over Feynman parameters,
which are relevant for the calculation of the virtual pair
correction to the muon decay spectrum ($z \gg m_e/m_\mu$ is
assumed).
\ba
\langle\frac{B}{D}\rangle &=& \frac{1}{3}\biggl[
\frac{1}{12}L_z^3 - \frac{5}{12}L_z^2
+ \biggl(\Li{2}{1-z} + \frac{14}{9} + \zeta(2)\biggr)L_z
- \Li{3}{1-z}
\nonumber \\
&+& 2\SSS{1,2}{1-z} - \frac{5}{3}\Li{2}{1-z}
+ \frac{1}{3} - \frac{10}{3}\zeta(2) - \zeta(3),
\biggr],
\nonumber \\
\langle\frac{yB}{D}\rangle &=& \frac{1}{3}\biggl[
\frac{1}{2}L_z^2 - \frac{8}{3}L_z + 2\Li{2}{1-z} + 2 + 4\zeta(2)
\biggr],
\nonumber \\
\langle\frac{yxB}{D}\rangle &=& \frac{z}{3(1-z)}\biggl[
- \ln z L_z - \Li{2}{1-z} + \ln^2z + \frac{8}{3}\ln z
\biggr],
\nonumber \\
\langle\frac{y^2B}{D}\rangle &=& \frac{1}{3}\biggl[
\frac{1}{4}L_z^2 - \frac{7}{12}L_z + \Li{2}{1-z}
- \frac{13}{2} + 5\zeta(2)
\biggr],
\nonumber \\
\langle\frac{y^2xB}{D}\rangle &=& \frac{z}{3(1-z)}\biggl[
- \frac{1}{2}\ln z L_z - \frac{1}{2}\Li{2}{1-z}
+ \frac{1}{2}\ln^2z + \frac{7}{12}\ln z
\biggr],
\nonumber \\
\langle\frac{y^2x^2B}{D}\rangle &=& \frac{z}{3(1-z)^2}\biggl[
\biggl( \frac{1-z}{2} + \frac{1}{2}z\ln z \biggr)L_z
+ \frac{z}{2}\Li{2}{1-z}
- \frac{z}{2}\ln^2z
\nonumber \\
&-& \frac{z}{12}\ln z - \ln z
- \frac{19}{2}(1-z)
\biggr],
\nonumber \\
\langle\ln\frac{D}{D_e}\rangle &=& \frac{1}{3(1-z)}
\biggl[  \frac{1-z}{4}L^2_z
+ \biggl( \frac{z}{2}\ln z - \ln z - \frac{19}{12}(1-z) \biggr)L_z
- \frac{z}{2}\Li{2}{1-z}
\nonumber \\
&-& \frac{z}{2}\ln^2z + \ln^2z + \frac{19}{12}(2-z)\ln z
- \frac{10}{3}(1-z) + 5(1-z)\zeta(2)
\biggr],
\nonumber \\
\langle\ln\frac{D}{D_\mu}\rangle &=& \frac{1}{3(1-z)}
\biggl[  \biggl( z - 1 - \frac{z}{2}\ln z \biggr)L_z
- \frac{z}{2}\Li{2}{1-z}
+ \frac{z}{2}\ln^2z
\nonumber \\
&-& \frac{5}{12}z\ln z + 2\ln z + \frac{19}{6}(1-z)
\biggr],
\\ \nonumber
L_z &\equiv& L + 2\ln z.
\ea
I used above a short notation for the integral over three
Feynman variables:
\ba
\langle F(v,x,y) \rangle =
\int\limits_0^1\frac{\dd v\;\phi(v)}{1-v^2}
\int\limits_{0}^{1}y\dd y\int\limits_{0}^{1}\dd x F(v,x,y),
\ea
and
\ba
D &=& y^2P_x^2 + (1-y)M^2, \qquad
P_x^2 = x^2m_\mu^2 + (1-x)^2m_e^2 + Bx(1-x), \qquad
\nonumber \\
D_\mu &=& y^2m_\mu^2 + (1-y)M^2, \qquad
D_e = y^2m_e^2 + (1-y)M^2,
\nonumber \\
B &=& zm_\mu^2\biggl(1+\frac{m_e^2}{m_\mu^2}\biggr), \qquad
M^2 = \frac{4m_e^2}{1-v^2}, \quad
\phi(v)=\frac{2}{3} - \frac{1}{3}(1-v^2)(2-v^2).
\ea

\section*{Appendix B\\
Asymptotic expressions for the muon form factor}

\setcounter{equation}{0}
\renewcommand{\theequation}{B.\arabic{equation}}

Using the Schwinger substitution~(\ref{schwinger}), I reproduced
the known~\cite{Burgers:1985qg,Hoang:1995ex} asymptotic
expressions for the $\order{\alpha^2}$ virtual pair contributions into the
Dirac form factor of muon:
\ba
&& F_1^{(4,a)}(m_1,m_2,Q^2)\bigg|_{m_1,m_2\ll Q^2} = \biggl(\frac{\alpha}{\pi}\biggr)^2
\biggl(\frac{e_2}{e}\biggr)^2 \biggl\{ - \frac{1}{36}L^3 + \frac{19}{72}L^2
\nonumber \\ && \qquad
- \biggl( \frac{265}{216} + \frac{\zeta(2)}{6} \biggr)L
+ D\biggl(\frac{m_1}{m_2}\biggr) \biggr\},
\\ \nonumber
&& D(1) = \frac{383}{108} - \frac{1}{4}\zeta(2),
\\ \nonumber
&& D(0)= \frac{3355}{1296} + \frac{19}{36}\zeta(2) - \frac{1}{3}\zeta(3), \\
&& D(R)\bigg|_{R \gg 1} = \frac{1}{36}l^3
- \frac{13}{72}l^2
+ \biggl( \frac{133}{216} + \frac{\zeta(2)}{3} \biggr)l
+ \frac{67}{54} - \frac{7}{36}\zeta(2) - \frac{1}{3}\zeta(3), \nonumber \\
\nonumber &&
L \equiv \ln\frac{Q^2}{m_2^2}\, , \qquad
l \equiv \ln R = \ln\frac{m_1^2}{m_2^2}\, ,
\ea
where $m_1=m_\mu$ is the muon mass; $m_2$ is the mass of the fermion in the loop;
$e$ and $e_2$ is the muon and fermion charges, respectively; $-Q^2$ is the square of
the momentum transferred in the spacelike region: $-Q^2 = (p_1 - p_2)^2 < 0$, where
$p_1$ and $p_2$ are the initial and the final muon four--momenta.


\end{document}